\font\caps=cmcsc10 at 12pt
\font\eightrm=cmr8 at 8pt
\documentclass[12pt]{oxarticle}
\usepackage{amsmath, amssymb}
\newcommand{\dB}{\d_{\rm BRST}}

\newcommand{\LT}{\LaTeX}

\newcommand{\bt}{\begin{tabular}{c}}
\newcommand{\et}{\end{tabular}}

\newcommand{\np}{\newpage} 
 
\newcommand{\eb}{\ee\be } 
\newcommand{\ebp}{\rt.\ee\be\lt.} 
\newcommand{\bmat}{\lt ( \begin{array} }
\newcommand{\emat}{  \end{array} \rt )}

\newcommand{\oH}{{\ov H}}
\newcommand{\oP}{{\ov P}}

\newcommand{\oQ}{{\ov Q}}

\newcommand{\ovD}{{\ov D}}

\newcommand{\cd}{{\cdot}}

\newcommand{\oq}{{\ov \q}}
\newcommand{\oT}{{\ov T}}

\newcommand{\ED}{\end{document}}

\newcommand{\og}{{\ov g}}
\newcommand{\oy}{{\ov \y}}

\newcommand{\oC}{{\ov C}}
\newcommand{\oL}{{\ov L}}

\newcommand{\A}{{\ov A}}

\renewcommand{\a}{\alpha}	
\renewcommand{\b}{\beta}

\renewcommand{\d}{\delta}
\newcommand{\e}{\epsilon}

\newcommand{\h}{\eta}
\newcommand{\q}{\theta}

\newcommand{\x}{\xi}

\newcommand{\y}{\psi}
\newcommand{\w}{\omega}

\renewcommand{\S}{\Sigma}

\newcommand{\la}{\label}
\newcommand{\ci}{\cite}

\newcommand{\ds}{\documentstyle}	
\newcommand{\fr}{\frac}

\newcommand{\pa}{\partial}
\newcommand{\ov}{\overline}
\newcommand{\be}{\begin{equation}}
\newcommand{\ee}{\end{equation}}
\newcommand{\ba}{\begin{array}} 
\newcommand{\ea}{\end{array}}
\newcommand{\bea}{\begin{eqnarray}}
\newcommand{\eea}{\end{eqnarray}}
\newcommand{\ra}{\rightarrow}
\newcommand{\Ra}{\Rightarrow}

\newcommand{\lt}{\left}
\newcommand{\rt}{\right}

\newcommand{\ben}{\begin{enumerate}}
\newcommand{\een}{\end{enumerate}}
\newcommand{\bitem}{\begin{itemize}}
\newcommand{\eitem}{\end{itemize}}

\newcommand{\articlenumber}{\LT 3232BreaksItselfLep}

\proofmodefalse
\begin{document}
\begin{center}

\vspace*{1in}
{\Huge  
Supersymmetry Breaks Itself \\for Quarks and Leptons\\
in the \\
SUSY Standard Model\\
[10pt] 
}

\renewcommand{\thefootnote}{\fnsymbol{footnote}}
%\footnotetext[1]{~here we have a footnote.}
\renewcommand{\thefootnote}{\arabic{footnote}}

{\caps John A. Dixon\footnote{cybersusy@gmail.com}\\Toronto, Canada}
\\[2in] 
%{\it No Fixed Address }
%\vfill
{\bf Abstract}
\end{center}

% This version is done in Miami, and it incorporates Margarets advice on 
%\LT 3213 and now it includes correction to \w \w term that I dropped by now understanding the J terms that survived, as set out in \LT 3225

%\Large
\small

Models like the Supersymmetric Standard Model (SSM)  possess simple, but well-hidden,  `Outfields'.  These Outfields are composite operators that violate superspace invariance, but in a special way.  For the Outfields, the violation is proportional to the field equations. So the Outfields can be, and are,   physically important. There is one such Outfield for every independent Lie algebra generator of invariance for the superpotential of the 
model.  These observations arise from a study of the BRST cohomology of such models. 

A new mechanism for SUSY breaking arises for a special  non-minimal version of the SSM, which will be called the CSSM.  The CSSM has right neutrinos and a Higgs singlet, which we call J, in addition to the usual SSM.  The superpotential for the CSSM has more symmetry than the SSM, and so the CSSM also has more Outfields.   For each Quark and Lepton in the CSSM, there exist  Outfields with the same Quark and Lepton quantum numbers, except that the Outfields have a dotted spinor index.
We form an extended Lagrangian, from the CCSM,  by coupling these Outfields to new `dotspinor superfields' to form the `Cybersusy Lagrangian'.  The resulting  action is supersymmetric,  but it is not superspace invariant.  The supersymmetry of this action depends on the invariances of the superpotential that gave rise to the Outfields.  

In the original CSSM, it is natural to add the simple linear superspace and gauge invariant term $m^2 J$.  This addition spontaneously breaks the gauge invariance $SU(2) \times U(1)$ down to $U(1)$. However, if this term is added to the  Cybersusy Lagrangian,  an interesting situation develops. The term $m^2 J$  is not invariant under the Lie Algebra that gave rise to the Quark and Lepton Outfields.  As a result, when $SU(2) \times U(1)$ spontaneously breaks down to $U(1)$, the supersymmetry for the Quarks and Leptons breaks at the same time.  The Cybersusy Lagrangian thus gives rise to a uniquely defined set of masses for the resulting Quark and Lepton spectrum after SUSY breaking.  This happens in much the same way that masses are generated in the Standard Model after the development of a VEV.  So the addition of the gauge invariant, and supersymmetric, term $m^2 J$ simultaneously breaks gauge invariance, and supersymmetry, for the Cybersusy Lagrangian.

This breaking of SUSY cannot be avoided, because it arises from  the local BRST cohomology of the theory, which is also the origin of the Outfields. It can also be seen that the Weak SU(2)  group, and the well-known remarkable  set  of doublets and  singlets for the Quarks, Leptons and Higgs,  have a {\em raison 
d'{\^e}tre} which relates to this mechanism.

 The SUSY breaking here   depends on only one parameter, which is the VEV that breaks $SU(2) \times U(1)$ to $U(1)$.   SUSY itself is not spontaneously broken here, so the vacuum energy remains zero after SUSY breaking. The resulting predictions for SUSY breaking are very constrained by the model.

\normalsize

\section{SUSY breaking in the SSM}

The Supersymmetric Standard Model (SSM) is a popular  method for extending the Standard Model to new physics, with an eye on the superstring as a unified theory . However the observed particles certainly do not occur in supermultiplets. So SUSY must  be a broken symmetry. 

\ben
\item
It is generally assumed that SUSY must be  `spontaneously broken'.  But there are serious problems
\ci{haber,martin,
SUSY06,SUSY07,SUSY08,SUSY09,SUSY10,weinbergvol3} which arise from spontaneous breaking of SUSY, including the following:   
\ben
\item
 The phenomenology arising from the spontaneous breaking of SUSY, in any model close to the SSM, has severe disagreement with experiment that arises from the mass sum rules
\ci{weinbergvol3,ferrarabook}. 
\la{sumruleprob}
\item
  This disagreement with experiment suggests that  the spontaneous breaking of SUSY must happen in a `hidden sector' so that the breaking has a chance of being realistic in the observable sector. 
\item
This  introduces many new problems that arise from `communicating the SUSY  breaking' to the observable particles, while avoiding the disagreement with experiment.
\item
There is also a problem with the   cosmological constant \ci{weinbergcosmoconstant,cosmoconstant}.  When SUSY is spontaneously broken, the  cosmological constant that arises  is naturally so   huge   that we could not possibly exist to concern ourselves about SUSY and its breaking.  This conclusion can be avoided if one accepts the   notion  that there is a fine tuning, for no obvious reason, to one part in $10^{50}$.
  This problem is not made less embarrassing by hiding it in a  `hidden sector'. Although we may not see the `hidden sector', gravity must react to it. Largely, this problem has been ignored while exploring phenomenology. 
\item
Another problem for SUSY is the notion that it is broken `dynamically' somehow.  This leads to tremendous calculational difficulties, since such a breaking  must be non-perturbative, and non-perturbative results are notoriously hard to derive in quantum field theory.   
\la{dynamprob}
\item
All known schemes for SUSY breaking generate such a large number of parameters, and have so little predictive power, that it has been stated \ci{Lykken}, partly as a joke, that `{Anything that is discovered at the LHC will be called Supersymmetry.}' . 
\la{lykken}
\een
\item
The construction of an effective theory of `explicitly softly broken' SUSY also has large problems, which include:
\ben
\item  There is no  clear way to interpolate between the effective theory and a fundamental theory of SUSY breaking. 
\la{clearwayprob}
\item
 The natural suppression of neutral flavour changing currents in the SM does not carry through to  the softly broken SSM, so the phenomenology tends to conflict with the experimental approximate conservation of flavour.
\la{flavourprob}
\item  The phenomenology tends to conflict with the approximate conservation of CP.
\la{cpprob}
\een
\een

\section{Cybersusy Made Simple--A New Action for SUSY breaking}

In this paper we shall construct an action and a new mechanism for SUSY breaking.  This SUSY breaking  is neither `spontaneous SUSY breaking' nor `explicit soft breaking'. Moreover this action arises out of the SSM itself.  One must calculate the BRST cohomology of the SSM to start this process, and that was done in \ci{jumps}, using the method of spectral sequences.

One of the attractive things about SUSY in 3+1 dimensions is Superspace 
 \ci{Superspace,WB,west,ferrarabook}. Superspace  is very useful for understanding the theory.  But it is  hard to imagine how SUSY manages to get broken, other than spontaneously, with the  strong confines of Superspace  to keep it unbroken.  However, in \ci{jumps} it was shown that BRST cohomology \ci{BRS,T,zinn}  provides us with a simple and unavoidable `escape from  Superspace'.  Certain `Outfield' operators arise\footnote{A generic example of an Outfield is in equation (\ref{Outfieldaction}) using the definition
 (\ref{psihat}). Outfields in the massless but interacting theory must satisfy the constraint (\ref{chiralconstraint}). }. They need to be written using  explicit factors of the Zinn sources, the supersymmetry ghosts $C, \oC$,  and the Superspace coordinates $\q, \oq$, in addition to superfields and Superspace derivatives. 
Some of these operators  have the quantum numbers of the Quarks and Leptons\footnote{The Leptonic examples are in equations 
(\ref{thetrickyoneforLeptonsdotspinor}), 
(\ref{rightelectdotspinor}) and (\ref{leftelectdotspinor}) 
 below.}. They also have very low dimension. 

In this paper we show  how to use these Outfields to write down a simple renormalizable  action  which conserves SUSY before the appearance of massive particles, but which breaks SUSY when  gauge symmetry breaks.  It could be argued, from looking at  the action, that we should interpret the Outfields as bound states in some sense, because they couple in a way that indicates they can decay into combinations of the Higgs and the usual Leptons. But perhaps this is the wrong interpretation, because, after all,  we do not view the Z boson as a bound state of muon and antimuon, although the action contains terms allowing that decay.

So it appears that the action here could be construed to  be an action which includes new multiplets,  and old multiplets, coupled in a way that is invariant under SUSY and gauge symmetry. The wonderful feature is that when gauge symmetry breaks, these mix in a way that violates SUSY. This arises because the couplings involve the Outfields, so that SUSY is `set up to break', because Superspace invariance is already broken, even though SUSY itself is not.  In fact the Lepton and Quark Outfields depend on peculiar invariances of the action\footnote{These invariances are written down in detail in \ci{jumps}. They are invariances of the massless superpotential only, and seem rather special to the CSSM.} that do not survive when gauge symmetry breaks.

SUSY breaking occurs in this `Cybersusy Action' at the same time as the spontaneous breaking of gauge symmetry, even though SUSY itself is not spontaneously broken or explicitly broken. It is `cohomologically broken'.   This SUSY breaking mechanism\footnote{The term `Cybersusy' was coined as an acronym standing for `{\bf C}ohomologicall{\bf Y} {\bf B}roken {\bf E}ffective {\bf R}etro SUSY, which is still appropriate for this Action.}  does not appear to suffer from  problems  (\ref{sumruleprob}) to  (\ref{clearwayprob}) in the above list, but the remaining two problems   (\ref{flavourprob}) and
(\ref{cpprob}) may still be an issue.
This paper arises out of the 
Cybersusy papers \ci{cybersusy} but this  is a much simpler and clearer version of Cybersusy, because now we can embody the ideas in a simple Lagrangian formalism.

Now we will construct the action.

\section{Superfields, Dotspinors, Outfields and the Construction of the Action in General}

We want to examine the CSSM, but it is easier to write down a generic action first, and then adapt it to the detailed notation needed for the CSSM in  section \ref{CSSMsection}.  So 
we start with the following action for a generic   theory of chiral superfields in 3+1 dimensions:
\[
{\cal A}_{\rm Chiral \;SUSY}
= \int d^4 x \;d^4 \q \lt \{
 {\widehat A}^i
{\widehat \A}_i \rt \}
\]\[
+ \int d^4 x \;d^2 \q \lt \{  \fr{1}{3}g_{ijk} {\widehat A}^i{\widehat A}^j
{\widehat A}^k + m^2 g_k {\widehat A}^k
+ {\widehat \Lambda}_k \d_{\rm SS}  {\widehat A}^k
\rt \}
\]\be
+ \int d^4 x \;d^2 {\ov \q} \lt \{  
\fr{1}{3} {\ov g}^{ijk} {\widehat \A}_i{\widehat \A}_j
{\widehat \A}_k
+ m^2 {\ov g}^k {\widehat \A}_k 
+ {\widehat {\ov \Lambda}}^k \d_{\rm SS}  {\widehat \A}_k
\rt \}
+ C^{\a} \oC^{\dot \b} Z_{\a \dot \b} 
\la{chiralsusyaction}
\ee
and add the following generic action for a chiral dotted spinor superfield (a `dotspinor' for short) in 3+1 dimensions:
\be
{\cal A}_{\rm Dotspinor}
= \int d^4 x \;d^4 \q  
 \; {\widehat \w}^{i \dot \b}
\pa_{\a \dot \b} {\widehat {\ov \w}}_{i}^{\a}
+ \int d^4 x \;d^2 \q \lt \{ m T_{ijk}{\widehat \w}^{i \dot \b} {\widehat \w}^{j}_{ \dot \b} {\widehat A}^k
+ {\widehat H}_{k \dot \a}	 \d_{\rm SS}
  {\widehat \w}^{k \dot \a} \rt \}+*
\la{dotspinoraction}
\ee
All our notation here was defined in \ci{jumps} and \ci{dotspinors}.   In this paper we  do not try to discuss the issues that arise from  gauged supersymmetry.  Anyway, the gauge sector does not affect the Leptonic sector for our purposes here, because the gauge theory does not carry Lepton number. 
  The terms ${\widehat \Lambda}_k $ and 
${\widehat H}_{k \dot \a}$ are Zinn sources for the SUSY variations.  Here the supersymmetry variation is $\d_{\rm SS}  =( C Q + \oC \oQ  +\x \pa) $. 
This action (\ref{dotspinoraction}) is quite peculiar.  Its equation of  motion has the form 
\be
\lt ( \Box^2 - m^4\rt ){\widehat \w}^{i \dot \b}=0
\ee
for the free theory.  However it does appear to make sense, in spite of this unusual equation of motion \ci{dotspinors}.

In order to find the Outfields, one starts with the nilpotent BRST cohomology operator $\dB$ of the above theory.  This is generated from the  BRST Poisson Bracket \ci{BRS,T,zinn}, which for this case has the form:
\[
{\cal P}_{ \rm BRST}[ {\cal A}]
=
\int d^4 x \; d^2 \q
\lt \{
\fr{\d {\cal A} }{\d {\widehat A}^i}  
 \fr{\d {\cal A} }{\d {\widehat \Lambda}_i}  
+\fr{\d {\cal A} }{\d {\widehat H}_i^{\dot \a}}  
 \fr{\d {\cal A} }{\d {\widehat \w}^i_{\dot \a}}  \rt \}
\]
\be
+\int d^4 x \; d^2 \oq
\lt \{
\fr{\d {\cal A} }{\d {\widehat \A}_i}  
 \fr{\d {\cal A} }{\d {\widehat {\ov \Lambda}}^i}  
+\fr{\d {\cal A} }{\d {\widehat \oH}^{i \a}}  
 \fr{\d {\cal A} }{\d {\widehat{\ov \w}}_{i\a}}  \rt \}
+
\fr{\pa {\cal A} }{\pa Z_{\a \dot \b} }  
 \fr{\pa {\cal A} }{\pa \x_{\a \dot \b}}  
\la{actionidentity}
\ee
The BRST operator is the `square root' of the BRST Poisson Bracket.  So starting with (\ref{actionidentity}), we have:
\[
{\d}_{ \rm BRST [ {\cal A}]}
\equiv
\int d^4 x \; d^2 \q
\lt \{
\fr{\d {\cal A} }{\d {\widehat A}^i}  
 \fr{\d  }{\d {\widehat \Lambda}_i}  
+ \fr{\d {\cal A} }{\d {\widehat \Lambda}_i}  
\fr{\d   }{\d {\widehat A}^i}  
+\fr{\d {\cal A} }{\d {\widehat H}_i^{\dot \a}}  
 \fr{\d   }{\d {\widehat \w}^i_{\dot \a}}   
+
 \fr{\d {\cal A} }{\d {\widehat \w}^i_{\dot \a}} 
 \fr{\d   }{\d {\widehat H}_i^{\dot \a}}   \rt \}
+\]
\[
\int d^4 x \; d^2 \oq
\lt \{
\fr{\d  {\cal A} }{\d {\widehat \A}_i}  
 \fr{\d   }{\d {\widehat {\ov \Lambda}}^i}  
+\fr{\d {\cal A} }{\d {\widehat {\ov \Lambda}}^i}  
\fr{\d   }{\d {\widehat \A}_i}  
+
\fr{\d {\cal A} }{\d {\widehat \oH}^{i \a}}  
 \fr{\d   }{\d {\widehat{\ov \w}}_{i\a}}   
+ \fr{\d {\cal A} }{\d {\widehat{\ov \w}}_{i\a}}
\fr{\d   }{\d {\widehat \oH}^{i \a}}  
  \rt \}
\]
\be
+
\fr{\pa {\cal A} }{\pa Z_{\a \dot \b} }  
 \fr{\pa   }{\pa \x_{\a \dot \b}}  
+
 \fr{\pa {\cal A} }{\pa \x_{\a \dot \b}}  
\fr{\pa  }{\pa Z_{\a \dot \b} }  
\la{BRSToperatorgeneric}
\ee
and it is easy to see that if (\ref{actionidentity}) is zero, then $\d_{ \rm BRST}$ is nilpotent: 
\be
{\cal P}_{ \rm BRST}[ {\cal A}]=0 \Ra 
\d_{ \rm BRST[ {\cal A}]}^2 =0
\la{nilpotenceinv}
\ee
If we take 
\be
{\cal A}_0 = {\cal A}_{\rm Chiral \;SUSY}
+{\cal A}_{\rm Dotspinor}
\ee
then the supersymmetry of this action is reflected in the identity:
\be
{\cal P}_{ \rm BRST}[ {\cal A}_0]
=0
\ee

In this paper, we want to consider the coupling of Outfields\footnote{The Outfields are discussed at length in \ci{jumps} and the present notation is introduced there.} to the dotspinors, and again we write this in a generic way:
\be
{\cal A}_{\rm Outfield\; Coupling}
=
 \int d^4 x \;d^2 \q \lt \{
 {\widehat \w}^{a\dot \a} 
{\widehat \oy}_{i \dot \a}
{\widehat A}^j  T_{aj}^{i}
\rt \}
+\int d^4 x \;d^2 \oq \lt \{ {\widehat {\ov \w}}_a^{  \a} 
{\widehat \y}^{i  \a}
{\widehat \A}_j  \oT_i^{aj}
\rt \}
\la{Outfieldaction}
\ee
In this action, we use the basic Outfield
 expressions \ci{jumps}:
\be
{\widehat \oy}_{i \dot \a}
=
\lt [
{\widehat {\Lambda}}_{i} \oC_{\dot \a} 
+ \ovD^2 \lt (
{\widehat \A}_{i} \oq_{\dot \a} 
\rt )
\rt ]
\la{psihat}\ee
and
\be
{\widehat \y}^{i  \a}
=
\lt [
{\widehat {\ov \Lambda}}^{i} \oC_{  \a} 
+ D^2 \lt (
{\widehat A}^{i} \q_{  \a} 
\rt )
\rt ]
\la{opsihat}
\ee
We shall assume for now that  $g_i m^2=0$ 
in (\ref{chiralsusyaction}). 
For this massless case, the Outfield constraint equation is: 
\be
T_{as}^{i} \og^{jks} \A_i \A_j \A_k =0
\la{chiralconstraint}
\ee
If (\ref{chiralconstraint}) is true, then it is a simple task  to verify that ${\cal A}_{\rm Outfield\; Coupling}$ satisfies the following equation:
\be
{\d}_{ \rm BRST [ {\cal A}_0]}{\cal A}_{\rm Outfield\; Coupling}=0
\la{incohom}
\ee
This was derived in \ci{jumps}, but we can verify it without all the machinery used there.  What happens is that the SUSY violation brought about by the explicit $\q$ parameters is compensated by the $\d {\widehat {\ov \Lambda}}^i = D^2 
{\widehat A}^i + \og^{ijk} {\widehat \A}_j {\widehat \A}_k $ field equation terms, provided that the trilinear terms 
(\ref{chiralconstraint}) cancel out. The point of
 (\ref{incohom}) is that the action 
(\ref{Outfieldaction}) is in the cohomology space of the operator ${\d}_{ \rm BRST [ {\cal A}_0]}$, even though  
(\ref{Outfieldaction}) explicitly depends on the Superspace parameters $\q,\oq$.  It takes some more work to show that this operator is not the boundary under $\dB$ of some other local expression, and the techniques of \ci{jumps} accomplish that in a general way.

The plan here is to build an action ${\cal A}$,  which incorporates the Outfield coupling (\ref{Outfieldaction}), and for which  the resulting new BRST Poisson Bracket is zero.  It is not immediately obvious that this can be done at all, because if we take:
\be
{\cal A}_1 = {\cal A}_{\rm Chiral \;SUSY}
+{\cal A}_{\rm Dotspinor}
+{\cal A}_{\rm Outfield\; Coupling}
\la{firstapproxgeneric}
\ee
we get
\be
 {\cal P}_{ \rm BRST}[ {\cal A}_1]
\neq 0
\ee
However, the failure of this equation is of 
order $\w \w$, and we shall see that it is possible to add terms to the action, order by order in $\w$, so that the BRST Poisson Bracket  of the final action is zero.  We shall calculate these terms below in detail, for the special case of the Leptons in the CSSM. They have the general form:
\be
{\cal A}_{\rm Outfield\; Completion}
=
\int d^4 x \;d^4 \oq {\widehat { A}}{\widehat {  \A}}\lt \{
  {\widehat {  \w}}\cd \oq 
 {\widehat { {\ov  \w}}}\cd \q + \cdots
+
{\widehat {  \w}}\cd \oq
{\widehat {  \w}}\cd \oq
{\widehat {  {\ov \w}}}\cd \q
{\widehat {  {\ov \w}}}\cd \q
\rt \} 
\la{genericcompletion}
\ee
Note that ${\widehat {  \w}}\cd \oq$ has dimension zero and that $\q^3 =0$, so that the above series stops as shown.
Now,  if we define the `Cybersusy Action' by: 
\be
{\cal A}  = {\cal A}_{\rm Cybersusy}  = {\cal A}_{\rm Chiral \;SUSY}
+{\cal A}_{\rm Dotspinor}
+{\cal A}_{\rm Outfield\; Coupling}
+{\cal A}_{\rm Outfield\; Completion}
\la{totalactiongeneric}
\ee
then we find that the relevant BRST Poisson Bracket equation 
\be
{\cal P}_{ \rm BRST}[ {\cal A}]
=0
\ee
is again true, which also implies 
 (\ref{nilpotenceinv}).  This then means that we have reinstalled supersymmetry into this action, while including the coupling to the Outfields and the new terms.  However this new action is clearly not a superspace invariant, because of the explicit $\q,\oq$ terms that it contains.

We also note that ${\cal A}_{\rm Cybersusy}$ defined by 
(\ref{totalactiongeneric}) is renormalizable, based on power counting, because there are no parameters with dimension of inverse mass in the action. This is as far as  we can go in this general and generic way. Now we must specify the fields and interactions that define the CSSM, and look at the details of the solutions for the Outfields in that case.

\section{Superfields, Dotspinors and Outfields  for the CSSM}

\la{CSSMsection}
Here is the Superspace potential for the  Supersymmetric Standard Model, modified to the CSSM:
\[
P_{{\rm CSSM}}   =
g \e_{ij} H^i K^j J
- g_{\rm J} m^2 J
+
p_{p q} \e_{ij} L^{p i} H^j P^{ q} 
\]
\be
+
r_{p q} \e_{ij} L^{p i} K^j R^{ q}
+
t_{p  q} \e_{ij} Q^{c p i} K^j T_c^{ q}
+
b_{p  q} \e_{ij} Q^{c p i} H^j B_c^{ q}
\la{CSSMsuperpotential}
\ee

The massless CSSM arises when we set $g_{\rm J} m^2=0$ in the Superpotential.    All of the above fields in  $P_{\rm CSSM} $ are the scalar parts of chiral superfields.    When   $g_{\rm J} m^2\neq 0$,  the $g_J$ term gives rise to the VEV which spontaneously breaks the gauge symmetry $SU(3) \times SU(2) \times U(1)$
 down to $SU(3) \times  U(1)$:

\be
g v^2 = g_J
\ee
\be
<H^i> = m h^i
\eb
<K^i> = m k^i
\ee
\be
h^i k_i = v^2
\ee
\be
 |h|=|k|=v
\ee

 We use ${\widehat E}^q$   to denote a superfield whose $\q,\oq$ independent part is the scalar $E^q$  . For example $L^{iq}$ is the scalar part of the left Leptonic SU(2) doublet chiral scalar superfield 
${\hat L}^{iq}$.  The spinor component is labelled $\y_{L \a}^{iq }$
and the auxiliary is $F_{L }^{iq}$.  The complex conjugate is 
${\widehat {\ov L}}_{iq}$.  The scalar component  of ${\widehat {\ov L}}_{iq}$ is labelled ${\ov L}_{iq}$, the spinor component  of ${\widehat {\ov L}}_{iq}$ is labelled $\oy_{L i q \dot \a}$
and the auxiliary is $\ov F_{L i q}$.  A chiral scalar superfield ${\widehat L}^{iq}$ satisfies the chiral constraint 
${\ov D}_{\dot \a} {\widehat L}^{iq}=0$, and its complex conjugate satisfies the antichiral constraint ${ D}_{ \a} {\widehat {\ov L}}_{iq}=0$.   Our notation is generally based on  that in \ci{Superspace}.

 After spontaneous breaking of gauge symmetry, the two $SU(2)$ doublets will give rise to the following superfields:
\be
Q^{c p i} \ra 
\lt (
\begin{array}{c}
U^{c p} 
\\
D^{c p} 
\\
\end{array}
\rt )
; L^{p i} \ra
\lt (
\begin{array}{c}
N^{p} 
\\
E^{p } 
\\
\end{array}
\rt )
\la{geghhgehhephtp}
\ee
The complex  matrices $p_{pq}$,  $r_{pq}$,  $t_{pq}$,  $b_{pq}$ are matched to, and named after, the right handed SU(2) singlet superfields ${\widehat P}^{q}$, ${\widehat R}^{q}$, ${\widehat T}^{cq}$,
 ${\widehat B}^{cq}$,  to form interaction terms, which also become mass terms when the gauge symmetry is spontaneously broken.
 Note that the second index of $p_{pq}$ is contracted with the right superfield ${\widehat P}^{q}$, etc.
Here is a table summarizing the various quantum numbers:
\be
\begin{tabular}{|c|c|c
|c|c|c
|c|c|c|}
\hline
\multicolumn{8}
{|c|}{Table (\ref{cssmtable}): The Chiral Superfields in the CSSM
}
\\
\hline
\multicolumn{8}
{|c|}{ \bf CSSM, Left  
Fields}
\\
\hline
{\rm Field} & Y 
& {\rm SU(3)} 
& {\rm SU(2)} 
& {\rm F} 
& {\rm B} 
& {\rm L} 
& {\rm D} 
\\
\hline
$ L^{pi} $& -1 
& 1 & 2 
& 3
& 0
& 1
& 1
\\
\hline
$ Q^{cpi} $ & $\fr{1}{3}$ 
& 
3 &
2 &
3 &
 $\fr{1}{3}$
& 0
& 1
\\\hline
$J$
& 0 
& 1
& 1
& 1
& 0
& 0
& 1
\\
\hline
\multicolumn{8}
{|c|}{ \bf CSSM, Right
Fields}
\\
\hline
$P^{ p}$ & 2 
& 1
& 1
& 3
& 0
& -1
& 1
\\
\hline
$R^{ p}$ & 0 
& 1
& 1
& 3
& 0
& -1
& 1
\\

\hline
$T_c^{ p}$ & $-\fr{4}{3}$ 
& ${\ov 3}$ &
1 &
3 &
 $-\fr{1}{3}$
& 0
& 1
\\
\hline
$B_c^{ p}$ & $\fr{2}{3}$ 
& ${\ov 3}$ 
&
1 &
3 &
 $- \fr{1}{3}$
& 0
& 1
\\
\hline
$H^i$ 
& -1 
& 1
& 2
& 1
& 0
& 0
& 1
\\
\hline
$K^i$ 
& 1 
& 1
& 2
& 1
& 0
& 0
& 1
\\
\hline
\end{tabular}
\\
\la{cssmtable}
\ee
%\vspace{.2cm}

In the above, Y is weak hypercharge, F stands for the number of families for each superfield, B is baryon number, L is Lepton number and D stands for mass dimension. 
Here is a table summarizing the various quantum numbers for the elementary chiral dotted spinor superfields that we will introduce:
\be
\begin{tabular}{|c|c|c
|c|c|c
|c|c|c|}
\hline
\multicolumn{8}
{|c|}{Table (\ref{dotspinortable}): The Quark and Lepton  Dotspinors  in  Cybersusy}
\\
\hline
\multicolumn{8}
{|c|}{ \bf Cybersusy Dotspinors, Right}
\\
\hline
{\rm Field} & Y 
& {\rm SU(3)} 
& {\rm SU(2)} 
& {\rm F} 
& {\rm B} 
& {\rm L} 
& {\rm D} 
\\
\hline
$ \w^{\dot \a}_{pi} $& +1 
& 1 & 2 
& 3
& 0
& -1
& $\fr{1}{2}$
\\
\hline
$ \w_{cpi}^{\dot \a} $ & $-\fr{1}{3}$ 
& 
${\ov 3}$ &
2 &
3 &
 $-\fr{1}{3}$
& 0
& $\fr{1}{2}$
\\
\hline
\multicolumn{8}
{|c|}{ \bf Cybersusy Dotspinors, Left}
\\
\hline
$\w_{E  p}^{\dot \a}$ &- 2 
& 1
& 1
& 3
& 0
& 1
& $\fr{1}{2}$
\\
\hline
$\w_{N  p}^{\dot \a}$ & 0 
& 1
& 1
& 3
& 0
& 1
& $\fr{1}{2}$
\\

\hline
$\w^{c \dot \a}_{U  p}$ & $+\fr{4}{3}$ 
& ${3}$ &
1 &
3 &
 $\fr{1}{3}$
& 0
& $\fr{1}{2}$
\\
\hline
$\w^{c \dot \a}_{D  p}$ & $-\fr{2}{3}$ 
& ${3}$ 
&
1 &
3 &
 $ \fr{1}{3}$
& 0
& $\fr{1}{2}$
\\
\hline
\end{tabular}
\\
\la{dotspinortable}
\ee
%\vspace{.2cm}

There are no dotspinors introduced in this Table that correspond to the Higgs/Goldstone superfields J,K,H.  These clearly mix with the gauge theory and so they cannot be considered without including the gauge theory.
The possible Yukawa or superpotential type interactions, 
following (\ref{dotspinoraction}), for the Leptonic sector of the CSSM are: 
\be
{\cal A}_{\rm  Dotspinor\; Yukawa} =
m \int d^4 x \;d^2 \q 
\lt \{ n^{pq} 
{\widehat \w}^{\dot \a}_{  pi}
{\widehat \w}_{N q \dot \a}
 {\widehat H}^i
+ e^{pq}
{\widehat \w}^{\dot \a}_{  pi}
{\widehat \w}_{E q \dot \a}
 {\widehat K}^i
 \rt \}+*
\la{dotspinoractioncssm}
\ee
We have taken Lepton number and SU(2) and hypercharge invariance into account here, and so these are the only possible terms with three fields.
There are no possible mass terms, before gauge symmetry breaking, in the CSSM, or in this action.  When gauge symmetry breaks, the above action yields masses in the usual way.

\section{Lepton Outfields and  Completion of the Action}

It was shown in \ci{jumps} that the 
following Outfields exist in the CSSM, and that they satisfy the constraint equations
 (\ref{chiralconstraint}):

{\bf Lepton Outfields}: {\em Chiral Dotted Spinor  Outfields with  
Quantum Numbers of the Leptons:}

\be
{\widehat \h}^{pi}_{L \dot \a}
=
g^{-1} {\widehat  L}^{p i} {\widehat  \oy}_{J\dot \a}
+
(p^{-1})^{qp} {\widehat K}^{ i}  {\widehat \oy}_{P q \dot \a}  
-
(r^{-1})^{qp} {\widehat H}^{ i} {\widehat \oy}_{R q \dot \a}   
\la{thetrickyoneforLeptonsdotspinor}
\ee
\be
{\widehat \h}^{p}_{P \dot \a}
=
g^{-1} {\widehat P}^{p } {\widehat  \oy}_{J\dot \a}
+
(p^{-1})^{pq} {\widehat K}^{ i} 
{\widehat \oy}_{L i q \dot \a}  
\la{rightelectdotspinor}
\ee
\be
{\widehat \h}^{p}_{R \dot \a}
=
g^{-1} {\widehat R}^{p }  {\widehat  \oy}_{J\dot \a}
-
(r^{-1})^{pq} {\widehat H}^{ i}  {\widehat \oy}_{L i q \dot \a}  
\la{leftelectdotspinor}
\ee
In this expression we use abbreviations like
(\ref{psihat}) and (\ref{opsihat}) for the basic Outfields:
\be
{\widehat \oy}_{L i q \dot \a}
=
\lt [
{\widehat {\Lambda}}_{Liq} \oC_{\dot \a} 
+ \ovD^2 \lt (
{\widehat {\ov L}}_{iq} \oq_{\dot \a} 
\rt )
\rt ]
\ee
and
\be
{\widehat  \oy}_{J\dot \a}
=
\lt [
{\widehat {\Lambda}}_{J} \oC_{\dot \a} 
+ \ovD^2 \lt (
{\widehat {\ov J}}  \oq_{\dot \a} 
\rt )
\rt ]
\ee

The interaction terms for these Outfields are:
\be
{\cal A}_{\rm Outfield\; Coupling}
=
\int d^4 x \; d^2 \q \;
\lt \{
{\widehat \w}^{\dot \a}_{  pi}
{\widehat \h}^{pi}_{L \dot \a}
+
{\widehat \w}^{\dot \a}_{N p }
{\widehat \h}^{p}_{R \dot \a}
+{\widehat \w}^{\dot \a}_{E p }
{\widehat \h}^{p}_{P \dot \a}
\rt \} + *
\la{leptonoutfieldcoupling}
\ee
Now we add this coupling between the  Outfields and the fundamental dotspinors in a first approximation to the full interacting action, just as we discussed at equation (\ref{firstapproxgeneric}) above:
\be
{\cal A}_1 = {\cal A}_{\rm Chiral \;SUSY}
+{\cal A}_{\rm Dotspinor}
+{\cal A}_{\rm Outfield\; Coupling}
\ee
Since we have a specific example here,  we can exhibit the completion terms (\ref{genericcompletion})
for this action in detail:

\be
{\cal A}_{\rm Outfield\; Completion}
=
{\cal A}_{\rm Double\; Dotspinor}
+
{\cal A}_{\rm Triple\; Dotspinor}
+
{\cal A}_{\rm Quadruple\; Dotspinor}
\la{completionaction}
\ee
where
\[
{\cal A}_{\rm Double\; Dotspinor}
=\int d^4 x d^4 \q\;
\lt \{
{\widehat {\ov \S}}^{ \a}_{0} \q_{\a}
{\widehat \S}^{\dot \a}_{0}
\oq_{\dot \a}
\rt .
\]
\be
+
{\widehat {\ov \S}}^{ \a}_{-iq} \q_{\a}
{\widehat \S}^{iq \dot \a}_{-}
\oq_{\dot \a}
+
{\widehat {\ov \S}}^{ \a}_{++q} \q_{\a}
{\widehat \S}^{q \dot \a}_{++}
\oq_{\dot \a}
+
\lt.
{\widehat {\ov \S}}^{ \a}_{0q} \q_{\a}
{\widehat \S}^{q \dot \a}_{0}
\oq_{\dot \a}
\ebp
+
{\widehat {\ov \w}}^{\a iq}
\q_{ \a} 
 \og^{-1} 
{\widehat {J} }  
{\widehat \S}^{iq \dot \a}_{-}
\oq_{\dot \a}
+
{\widehat \w}^{\dot \a}_{iq}
\oq_{\dot \a} 
 g^{-1} 
{\widehat {\ov  J} }  
{\widehat {\ov \S}}^{ \a}_{-iq} \q_{\a}
\rt \}
\la{doubledotstuff}
\ee

where
\be
{\widehat \S}^{\dot \a}_{0}
=
g^{-1}
\lt \{
{\widehat \w}^{\dot \a}_{E p }
 {\widehat P}^{p } 
 +
{\widehat \w}^{\dot \a}_{N p }
 {\widehat R}^{p } 
+{\widehat \w}^{\dot \a}_{ pi}
 {\widehat  L}^{p i} 
\rt \}
\la{termsfromJ}
\ee
\be
{\widehat \S}^{iq \dot \a}_{-}
=
\lt \{
{\widehat \w}^{\dot \a}_{E p }
(p^{-1})^{pq}      {\widehat K}^{ i} 
+
{\widehat \w}^{\dot \a}_{N p }
(r^{-1})^{pq}      {\widehat H}^{ i} 
\rt \}
\la{termsfromL}
\ee
\be
{\widehat {\ov \S}}^{ \a}_{++q} \q_{\a}
= 
{\widehat \w}^{\dot \a}_{ pi}
 (p^{-1})^{qp} {\widehat K}^{ i} 
\la{termsfromP}
\ee
\be
{\widehat \S}^{q \dot \a}_{0}
=
{\widehat \w}^{\dot \a}_{ pi}
\oq_{\dot \a} (r^{-1})^{qp} {\widehat H}^{ i} 
\la{termsfromR}
\ee
and
\be
{\cal A}_{\rm Triple\; Dotspinor}
= 
\int d^4 x d^4 \q
\lt \{
{\widehat {\ov\w}}^{iq  \a }
\q_{\a}
\og^{-1}  
 {\widehat {\ov \S}}_{- iq}^{  \b} \q_{ \b}
{\widehat \S}_0^{\dot \a} \oq_{\dot\a}  
+*
\rt \}
\la{tripledotstuff}\ee
and
\be
{\cal A}_{\rm Quadruple\; Dotspinor}
=
\int d^4 x d^4 \q
\lt \{
{\widehat {\ov\w}}^{iq  \a }
\q_{\a}
\og^{-1}  
 {\widehat {\ov \S}}_{- iq}^{  \b} \q_{ \b}
{\widehat \S}_-^{jp\dot \a} \oq_{\dot\a}  
{\widehat {\w}}_{jp}^{ \dot \b }
\oq_{\dot \b} g^{-1} 
\rt \}
\la{quadrupledotstuff}\ee

\section{The Cybersusy Action  for Quarks and Leptons based on the CSSM}

We now have constructed an action that consists of terms of the form
(\ref{totalactiongeneric}), except that we must translate the generic form into the CSSM form using the above information regarding the field content, the  superpotential, the Outfields, the dotspinor Yukawa terms,  and the completion terms.  Similarly we could write down the detailed form of the BRST operator $\dB$ in (\ref{BRSToperatorgeneric}) in terms of the CSSM superfields and dotspinors, and we needed to do this to find the form of (\ref{completionaction}).  We will not write out these rather long expressions explicitly here.
But here is how the action needs to be assembled, in a summary form.  In the above,  we have written only the Lepton forms, but the Quark forms are identical, except for a change of notation and  the addition of colour indices.  There are four major terms:

\[
 {\cal A}_{\rm Cybersusy}= \]
\be
{\cal A}_{\rm Chiral \;SUSY}
+{\cal A}_{\rm Dotspinor}
+{\cal A}_{\rm Outfield\; Coupling}
+{\cal A}_{\rm Outfield\; Completion}
\la{totalactioncybersusy}
\ee
In the above,
\ben
\item
The subaction ${\cal A}_{\rm Chiral \;SUSY}
$ is of the form in (\ref{chiralsusyaction}), except that the fields ${\widehat A}^i$ must be replaced by
the fields in Table (\ref{cssmtable}), and the superpotential must be replaced by the form 
(\ref{CSSMsuperpotential}).  We set the terms  $g_J m^2 =0$ for the massless case, which is used to construct the Cybersusy action above.
\item
The subaction ${\cal A}_{\rm Dotspinor}$ is of the form in (\ref{dotspinoraction}), except that the fields ${\widehat \w}^i_{\dot \a}$ must be replaced by
the  fields in Table (\ref{dotspinortable}). Note that one needs to add the specific form of ${\cal A}_{\rm  Dotspinor\; Yukawa}$ in equation
(\ref{dotspinoractioncssm}), and similar terms for the Quarks.
\item
The subaction ${\cal A}_{\rm Outfield\; Coupling}
$ has the form  (\ref{leptonoutfieldcoupling}), but one also needs to add a Quark version of this too.
\item
The subaction ${\cal A}_{\rm Outfield\; Completion}$ is of the form 
(\ref{completionaction}), which consists of the three kinds of terms listed below equation (\ref{completionaction}), namely (\ref{doubledotstuff}),
  (\ref{tripledotstuff}) and 
(\ref{quadrupledotstuff}).
  One also needs to add the Quark version of these. 
\item
The result is the Cybersusy Action ${\cal A}\equiv {\cal A}_{\rm Cybersusy}$ for the Quarks and Leptons based on the CSSM. 
It satisfies the BRST Poisson Bracket (\ref{nilpotenceinv}),  which summarizes SUSY invariance:
\be
{\cal P}_{ \rm BRST}[ {\cal A}]=0 
\la{brstforcyber}
\ee
However, ${\cal A}\equiv {\cal A}_{\rm Cybersusy}$  contains plenty of $\q,\oq$ in addition to superfields and superspace derivatives, so it is not a superspace invariant.  It is not manifestly supersymmetric. But it is supersymmetry invariant in the sense of 
(\ref{brstforcyber}), and to achieve this it uses  the  subtle features that arise from the existence of the Outfields for the Quarks and Leptons in the CSSM. 
\item
 However this identity (\ref{brstforcyber}) is true only  for the massless case where gauge symmetry is not yet broken.  When $g_J m^2 \neq 0$,   gauge symmetry breaking occurs, and SUSY breaking follows. The behaviour of $
 {\cal A}_{\rm Cybersusy}$ for that case  is discussed below in section \ref{breakcssmsec}. 

\een

\section{SUSY Breaking for Quarks and Leptons}

\la{breakcssmsec}

For convenience we shall focus only on the Leptons.  The Quarks follow identically except for the change of notation and the addition of color indices.
From \ci{jumps} we know that the Outfields do not survive in the cohomology space of $\dB$ when the gauge symmetry breaks in the CSSM.  This means that  we  get the following for ${\cal A}  \equiv {\cal A}_{\rm Cybersusy}$  when the VEV appears:
\be
 {\cal P}_{ \rm BRST}[ {\cal A}]
=
\int d^4 x \;d^2 \q 
g_J m^2 \lt \{  
{\widehat \w}_{  pi}^{\dot \a}
{\widehat L}^{pi} \oC_{\dot \a}
+
{\widehat \w}_{E p}^{ \dot \a}
{\widehat P}^{p} \oC_{\dot \a}
+
{\widehat \w}_{N p}^{ \dot \a}
{\widehat R}^{p} \oC_{\dot \a}
 \rt \}+*
\eb+ {\rm More\; terms \;from \;{\cal A}_{\rm Outfield\; Completion}}
\la{lookslikeananomaly}
\ee
Moreover, the local BRST cohomology tells us that we cannot find any local terms which will compensate this result \ci{jumps}.  The BRST Poisson Bracket cannot be restored when the VEV is present in the CSSM.
This means that SUSY is broken, and the breaking is proportional to the constant $g_J m^2$, which is what gives rise to the VEV. 

So we expect to find that the mass spectrum of the theory is not supersymmetric if $g_J m^2 \neq 0$.  
Let us look at the kinetic and mass quadratic terms that arise for Leptons in the action.  Here are the quadratic mixing terms from the Outfield action:
\[
{\cal A}_{\rm Outfield\; Quadratic \;Mixing}
=
\int d^4 x \; d^4 \q \;
\lt \{
{\widehat \w}^{\dot \a}_{  pi}
\lt ( 
(p^{-1})^{qp} m k^{ i}  {\widehat \oP}_{ q}   \oq_{ \dot \a}  
-
(r^{-1})^{qp} m h^{ i} {\widehat {\ov R}}_{ q} \oq_{ \dot \a } 
 \rt )
\rt.
\]
\be
\lt.+
{\widehat \w}^{\dot \a}_{E p }
(p^{-1})^{pq} m k^{ i} 
{\widehat \oL}_{ i q} \oq_{ \dot \a}  
+{\widehat \w}^{\dot \a}_{N p }
(r^{-1})^{pq} m h^{ i}  {\widehat \oL}_{i q } \oq_{\dot \a}  
\rt \}
\ee
The appearance of the VEV has caused these trilinear interactions  to give rise to bilinear mixing terms. 
There are also quadratic mixing terms which appear from
 (\ref{doubledotstuff}), due to the fact that the expressions
 (\ref{termsfromL}), 
 (\ref{termsfromP}), and
 (\ref{termsfromR}) 
become  linear terms when the fields $H,K$ take their VEVs. 
The resulting terms are:
\be
{\cal A}_{\rm Outfield\; Completion\; Quadratic \;Mixing\; Terms}
\eb
=\int d^4 x d^4 \q\; m^2 
\lt \{
{\widehat {\ov \S}}^{ \a}_{-iq\; {\rm VEV}} \q_{\a}
{\widehat \S}^{iq \dot \a}_{-\; {\rm VEV}}
\oq_{\dot \a}
\ebp
+
{\widehat {\ov \S}}^{ \a}_{++q\; {\rm VEV}} \q_{\a}
{\widehat \S}^{q \dot \a}_{++\; {\rm VEV}}
\oq_{\dot \a}
+
{\widehat {\ov \S}}^{ \a}_{0q\; {\rm VEV}} \q_{\a}
{\widehat \S}^{q \dot \a}_{0\; {\rm VEV}}
\oq_{\dot \a}
\rt \}
\la{doubledotstuff2}
\ee
where

\be
{\widehat \S}^{iq \dot \a}_{-\; {\rm VEV}}
=
\lt \{
{\widehat \w}^{\dot \a}_{E p }
(p^{-1})^{pq}      {  k}^{ i} 
+
{\widehat \w}^{\dot \a}_{N p }
(r^{-1})^{pq}      {  h}^{ i} 
\rt \}
\la{termsfromLvev}
\ee
\be
{\widehat {\S}}^{q \a}_{++\; {\rm VEV}}  
= 
{\widehat \w}^{\dot \a}_{ pi}
 (p^{-1})^{qp} {  k}^{ i} 
\la{termsfromPvev}
\ee
\be
{\widehat \S}^{q \dot \a}_{0\; {\rm VEV}}
=
{\widehat \w}^{\dot \a}_{ pi}
 (r^{-1})^{qp} {k}^{ i} 
\la{termsfromRvev}
\ee

These two sets of mixing terms are to be added to the usual kinetic terms and mass terms that arise from 
(\ref{chiralsusyaction})
and (\ref{dotspinoraction}).  Then the masses are determined by the positions of the poles in the propagators, which are the inverse of these mixed-up mass and kinetic terms. The Quarks work exactly the same way as the Leptons, as one can see from the fact that their Outfields are exactly like the Lepton Outfields, except for the presence of the SU(3) index. This mixing problem for the Leptons will not be further considered in this paper except to say:
\ben
\item
If Cybersusy makes sense, then the mass spectrum needs to  make sense, which means that
\ben 
\item
Values of ${\rm mass}^2$ should not be negative.
\item
The known Leptons and Quarks
should have the lowest mass for a reasonable range of the parameters
\item
No strange problems like unitarity violation or `ghosts' should emerge.
\een
\item
A very similar problem was solved in the first papers on Cybersusy \ci{cybersusyIV}, and the spectrum there did appear to have the above properties.  This calculation needs to be done again with the present simpler presentation.  The two formulations appear to be dual in some sense, and the spectra might be equivalent. 
\item
Because the  SUSY breaking arises from the one parameter that yields the VEV, the spectrum must be quite constrained, and a SUSY signature for experiment  might  be fairly clear. 
\een

\section{Supergravity, SUSY Anomalies and Unitarity}

\la{unitaritysection}

Since the equation (\ref{lookslikeananomaly})
looks as if a supersymmetry anomaly is present, it is natural to be concerned that this mechanism will interfere with the unitarity of supergravity. That would indeed be a concern,  except for a curious fact.  The fact is that  chiral dotted spinor superfields cannot be coupled to supergravity \ci{westpage138,Superspace,WB}. So we are left with a puzzle here, but perhaps this actually makes some sense. Here is a possible resolution, but there are serious issues that arise here:
\ben
\item
The   Cybersusy Action (\ref{totalactiongeneric}) is an effective action,  to be used only for the rigid SUSY theory, and only for energies that are low compared to the mass scale of  supergravity, which might be, say, the Planck mass of $10^{19}$ Gev.
\item
The action that couples to supergravity is the usual CSSM
without any dotspinor superfields. 
\item
So perhaps  we are   forced to a conclusion of the following kind:
\ben
\item
Supergravity does not have a massive gravitino.  In  effect, SUSY does not need to be broken, as far as supergravity is concerned. 
\item
There is no contribution to the cosmological constant from supersymmetry breaking \ci{cosmoconstant}, because there is no spontaneous breaking of SUSY, there is only spontaneous breaking of gauge symmetry, which leaves the vacuum with zero energy.
\item
There is no problem with unitarity for supergravity, because supergravity is `blind' to the breaking of SUSY and it cannot `see' the anomalous behaviour in (\ref{lookslikeananomaly}).
\item
Supergravity is coupled only to the elementary particles, and it is not directly coupled to their bound states in addition.
But this begs the question of what `elementary' means.  Perhaps that is not even a sensible question in this context.
\een
\een

\section{Conclusion: The Dotspinors Break SUSY when the VEV Arises}

We do not need to know the answers to the hard questions in section \ref{unitaritysection} for what  we want to do here.  There are more practical, and easier, issues that can be examined.  If these practical issues do not kill Cybersusy somehow, then the hard questions in section \ref{unitaritysection} may require answers.  Let us summarize the situation:
\ben
\item
 We have found a uniquely defined supersymmetric action in (\ref{totalactioncybersusy}), which is apparently renormalizable, both before and after SUSY breaking.
\item
This action (\ref{totalactioncybersusy}) includes the CSSM as a subaction.  To see this one simply sets all the  dotspinors (and their Zinns) to zero.  
\item
The supersymmetry of  (\ref{totalactioncybersusy})  is very unstable, partly because it is already violating superspace invariance through the Outfields, and partly because the Outfields do not stay in the Cohomology Space of the CSSM when the VEV appears. 
\item
 As soon as the VEV  breaks gauge symmetry, the SUSY gets broken too, but the breaking of SUSY is not of the  spontaneous variety.  
\item
Before gauge symmetry breaking, all the particles are massless and supersymmetric. 
\item
All the Leptons are massive after the VEV appears,  but the Quarks and Leptons do not lie in Supermultiplets of mass.
\item
 The theory has a well defined spectrum for broken SUSY, depending on only one SUSY breaking parameter (the VEV), plus a number of SUSY conserving parameters in the interactions.  
\item
The breaking of SUSY in the CSSM arises from the dotspinors, which mix with the ordinary Leptons and Quarks, and each other, in a non-supersymmetric way after the VEV arises.
Clearly, the SUSY breaking  comes from the Outfields and their coupling to the new dotspinors. 
\item
If one does not couple the new dotspinors in this way, the SUSY of the CSSM is undisturbed by the VEV.

\item
 This mechanism for SUSY breaking is very constrained.
\een

So the first question to answer is whether the spectrum makes sense, and whether it is consistent with the experimental result, which is that the lowest mass Leptons have spin $\fr{1}{2}$, and that any superpartners must be very massive. That is relatively easy, but also a little complicated, to calculate with the above results.  It can be done with only one flavour of course, and that is the simplest way to start.

 It is also simple to couple this theory to the gauge theory \ci{dotspinors}.  We have only looked at Leptons in this paper. Quarks work the same way.  Assuming that the spectrum is not a problem, the next major hurdle will be the breaking of SUSY in the Higgs/Goldstone/Gauge sectors. Preliminary results, using the same sorts of ideas, are encouraging, so far.  But it seems probable that there are already interesting signatures for SUSY that can be extracted from the above action. One notable prediction is that there should be a heavy vector boson lepton.  

\vspace{.2in}

\np
\hspace{2.5in} {\bf Acknowledgments}

I thank Philip Candelas,  Brian Batell,  Cliff Burgess, Peter Scharbach, J.C. Taylor and  Peter West  for useful  correspondence, conversations and remarks.

%\end{document}

\tableofcontents

 {\tiny \articlenumber}

\begin{thebibliography}{99}
\bibitem{cybersusy} Ibid.,  `Cybersusy'-- The author has written a number of papers on this subject.  These papers were generated during  attempts to verify, understand and apply the complicated mathematical results in \ci{jumps}, and in particular they arose from the cohomology of the CSSM. Those papers placed the cart before the horse, since \ci{jumps} was still in draft form at the time they were written.  There was plenty of confusion in those papers, much of which is now resolved in this paper.  

The major change is that there is now a Lagrangian for Cybersusy.  This radically improves the phenomenological utility of Cybersusy, provided that the spectrum does make sense, and that there are no other serious problems.  The Lagrangian here contains separate fields for the dotspinors, coupled to the Outfields, whereas in the Cybersusy papers, this was not done because  the Superspace version of the Outfields was not understood  then, and it looked impossible to couple the dotspinors to the Outfields. 


The early Cybersusy papers are 
 \ci{cybersusyI,cybersusyII,cybersusyIII,cybersusyIV,
cybersusyV}, and most of the useful parts of these are all now contained in the present short paper, except for the spectrum calculation in \ci{cybersusyIV}. The paper \ci{susy09} is a brief summary of those first five papers, and it contains some of the confusion in them, which is now resolved here. 
All the useful parts of the paper \ci{symmetricterms} are now  included in \ci{jumps}. The papers  \ci{dotspinors} and \ci{cosmoconstant} are still  relevant, and will soon be brought up to date using the results here and in \ci{jumps}.
\end{thebibliography}
\end{document}